\documentstyle[fleqn,espcrc2,epsf]{article}
\title{
$I=2$ Pion Scattering Length from Two-Pion Wave Function
\thanks{presented by N. Ishizuka}
}
\author{
N.~Ishizuka
\address{
Center for Computational Physics,
University of Tsukuba,Tsukuba, Ibaraki 305-8577, Japan
}$^{\rm ,b }$
and
T.~Yamazaki
\address{
Institute of Physics,
University of Tsukuba,
Tsukuba, Ibaraki 305-8571, Japan
}
}
\begin{document}
%
%
\begin{abstract}
We present a preliminary report on a calculation of 
scattering length for $I=2$ $S$-wave two-pion system 
directly from the two-pion wave function.
Results are compared with those calculated from 
the time dependence of two-pion four-point functions.  
Calculations are made with an RG-improved action for gluons and improved 
Wilson action for quarks
at $a^{-1}=1.207(12)$~GeV on
$20^3 \times 48$ and 
$24^3 \times 48$ lattices. 
\end{abstract}
\maketitle
%
%
\section{ Introduction }
By now the standard procedure employed for calculating
the scattering length of hadrons from Euclidean lattice QCD is to 
use the finite-size method proposed by L\"uscher, 
which relates the finite volume shift of energy eigenvalues $\Delta E$ 
to the scattering lengths. 
The formula reads~\cite{Luscher}
\begin{equation}
a_0/(L\pi) = - x - A\cdot x^2 - B\cdot x^3 + O(x^4)
\label{Luscher.eq}
\end{equation}
where $x = \Delta E \cdot  2 m_\pi L^2 / (4\pi)^2$, and 
$A = -8.9136\dots$ and $B = 95.985\dots$ are geometrical constants. 
The energy shift $\Delta E$ is extracted from the time dependence of 
two-pion four-point functions.  
In this way a number of calculations has been 
made for the $I=2$ $S$-wave two-pion scattering length
using the staggered~\cite{SGK,Kuramashi},  
the standard~\cite{Kuramashi,GPS,CP-PACS.SCL,JLQCD} and 
the improved Wilson fermion actions~\cite{LZCM}.

It should be noted that
the condition $R < L/2$ is assumed
for the two-pion interaction range $R$ and the lattice volume $L^3$
in derivation of the formula (\ref{Luscher.eq})~\cite{Luscher}.
So far there have been 
studies of the lattice volume dependence of the scattering length, 
but no direct investigation of the interaction range $R$.
It is extremely important that 
we examine the validity of the necessary condition for the L\"uscher's formula 
in our current lattice simulations.
This can be done by investigating the two-pion wave functions, 
which is one of the purpose of this article.
Similar work has been carried for 2 dimensional XY model~\cite{Wisz.stat}.

Once one has access to the wave functions, 
one can try to extract the scattering length from them.  
This is the second, and perhaps more interesting, 
purpose of this article. 
%
%
\section{ Method }
Our idea is based on the derivation of the L\"uscher's formula 
(\ref{Luscher.eq}).
L\"uscher found that the two-pion wave function $\phi(\vec{x})$
on a finite periodic box $L^3$
satisfies the following effective Schr\"odinger equation~\cite{Luscher}.
\begin{equation}
  ( \triangle + k^2 ) \phi( \vec{x} ) 
  = \int {\rm d}^3 y \ U_{k} (\vec{x},\vec{y}) \phi(\vec{y})
\label{Luscher.two}
\end{equation}
where $\vec{x}$ is the relative coordinate of the two pion and
$k^2$ is related to the two-pion energy eigenvalue 
through $E=2\cdot\sqrt{ m_\pi^2 + k^2 }$.
The function $U_k (\vec{x},\vec{y})$ is the Fourier transform of 
the modified Bethe-Salpeter kernel introduced in \cite{Luscher}.
It is non-local and generally depends on energy.

In the derivation of the formula
it is assumed that $U_k ( \vec{x} , \vec{y} )\sim 0$ for $R < |\vec{x}| , |\vec{y}|$.
In this region the wave function satisfies the Helmholtz equation 
$( \triangle + k^2 ) \phi( \vec{x} ) = 0$.
The general solution of Helmholtz equation on a finite periodic box $L^3$
can be written as
\begin{equation}
  \phi( \vec{x} ) = \sum_{\vec{n} \in {\cal Z}^3} 
                       {\rm e}^{ i \vec{x} \cdot \vec{n} /  N }
                     / \bigl( n^2 - k^2\cdot N^2 \bigr)
\label{Luscher.div_two}
\end{equation}
up to over all constant, where $N=L/(2\pi)$.
Assuming the interaction range is smaller than one half of box size, 
{\it ie.} $R < L/2$, 
and expanding the general solution (\ref{Luscher.div_two}) in terms of 
spherical Bessel $j_l(x)$ and Noeman $n_l (x)$ functions 
for $R < |\vec{x}| < L/2$, we obtain
\begin{equation}
  \phi( \vec{x} ) = 
    \alpha_0 ( k ) \cdot j_0 ( k |\vec{x}| ) + \beta_0 ( k ) \cdot n_0 ( k |\vec{x}| ) 
  + \cdots
\label{Luscher.div_three}
\end{equation}
where neglected terms are contributions from states with angular momentum 
$l \geq 4$.
The expansion coefficients $\alpha_0 ( k )$ and $\beta_0 ( k )$ 
yield the scattering phase shift in infinite volume by 
$\tan \delta_0 ( k ) = \beta_0 ( k ) / \alpha_0 ( k )$.
In particular for the lowest energy state of the two-pion,
it gives the scattering length by
$\beta_0 ( k ) / \alpha_0 ( k ) = a_0 \cdot k + {\rm O}(k^2)$.
The constant $\beta_0 ( k )$ are also 
related to $\alpha_0 ( k )$ geometrically 
and the relation leads to the L\"uscher's formula (\ref{Luscher.eq}).
%
%
\begin{figure}[t]
\vspace*{-0.2cm}
\centerline{\epsfxsize=7.2cm \epsfbox{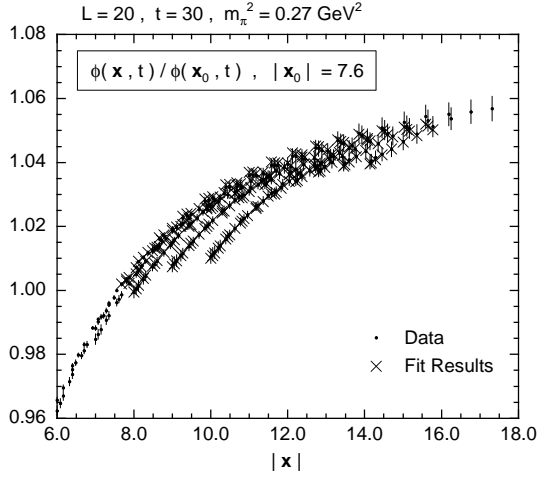}}
\vspace*{-1.0cm}
\caption{
\label{FIG.1.fig}
Two-pion wave function on $20^3$ lattice at $t=30$ and $m_\pi^2=0.27{\rm GeV}^2$.
}
\vspace*{-0.5cm}
\end{figure}
%
%

In this work we define the two-pion wave function by 
\begin{equation}
     \phi( \vec{x}, t ) = \sum_{ {\bf R} , \vec{X}} 
     \bigl\langle \ \pi( R(\vec{x}) + \vec{X},t) \ \pi(\vec{X},t)
     \ \ S  \ \bigr\rangle
\label{def_of_ppwf}
\end{equation}
where ${\bf R}$ is an element of cubic group, and 
summation over ${\bf R}$ and $\vec{X}$ 
projects out the ${\bf A^{+}}$ sector of the cubic group and that of  
the zero center of mass momentum.
In order to enhance signals
we use a source constructed with two wall sources given by 
$S=W(t_1) W(t_2)$, with $W(t)$ the wall source at $t$,  
in Coulomb gauge fixed configurations.
%
%
\section{ Results }
We work in quenched lattice QCD employing an RG-improved action for gluons at 
$\beta=2.334$ 
and an improved Wilson action for quarks at $C_{SW}=1.398$.
The corresponding lattice cutoff is estimated as $1/a=1.207(12)$~GeV from $m_\rho$.
The volumes of lattices (number of configuration) are
$20^3\times 48$ $(150)$ and
$24^3\times 48$ $(100)$ which correspond to 
$3.26{\rm fm}^3$ and 
$3.92{\rm fm}^3$ in physical units.
Quark masses are chosen to be
$m_\pi^2 ( {\rm GeV}^2 ) = 0.27$, $0.44$, and $0.74$.
Quark propagators are solved with 
the Dirichlet boundary condition imposed in the time direction 
and the periodic boundary condition in the space directions. 
The sources are set at $t_1=7$ and $t_2=8$ for both lattices.

In Figure~\ref{FIG.1.fig}
we show the ratio of the wave function $\phi( \vec{x}, t )/\phi( \vec{x}_0, t )$
on a $20^3$ lattice at $t=30$ and $m_\pi^2=0.27 {\rm GeV}^2$. 
Here we choose $|\vec{x}_0|=7.6$ for normalization point.
We find that the statistical error is very small.
%
%
\begin{figure}[t]
\vspace*{-0.2cm}
\centerline{\epsfxsize=7.0cm \epsfbox{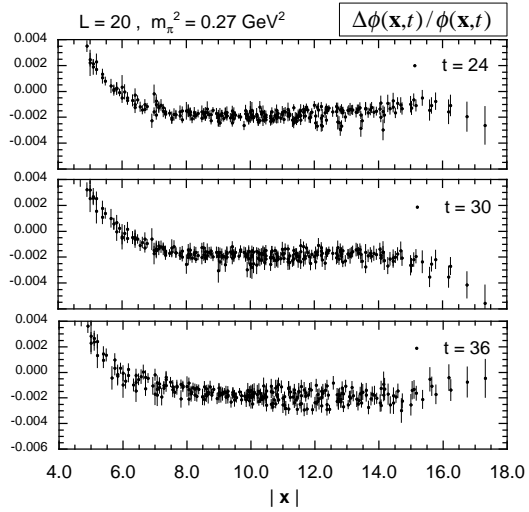}}
\vspace*{-1.0cm}
\caption{\label{FIG.2.fig}
$\triangle \phi(\vec{x},t) / \phi(\vec{x},t )\ ( \sim - k^2)$
on $20^3$ lattice at $m_\pi^2=0.27 {\rm GeV}^2$
for several $t$.
}
\vspace*{-0.5cm}
\end{figure}
%
%
%
Figure~\ref{FIG.2.fig} shows 
the ratio $\triangle \phi(\vec{x},t)/\phi(\vec{x},t)$ 
at the same parameters for several $t$.
This ratio should take a negative constant $-k^2$
for the $I=2$ two-pion system for $R < |\vec{x}|$.
Clear plateau appears for 
$|\vec{x}| > 7.5$ at $t \geq 30$ as shown in the figure.
This indicates that 
the necessary size for application of the L\"uscher's formula is 
$L > 2 \times 7.5 a\ ( = 2.45 {\rm fm}$ ).

\begin{figure}[t]
\vspace*{-0.4cm}
\centerline{\epsfxsize=7.2cm \epsfbox{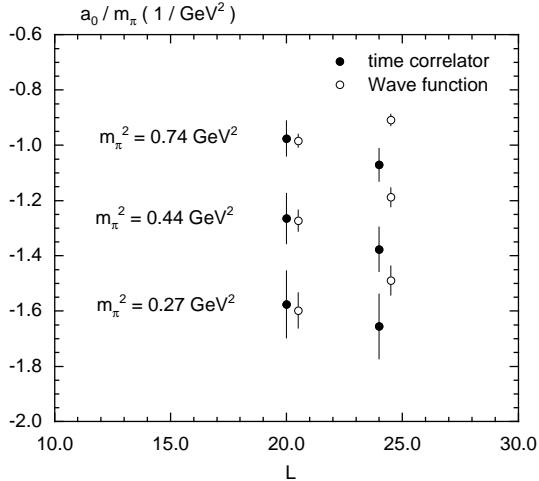}}
\vspace*{-1.0cm}
\caption{\label{FIG.3.fig}
$I=2$ two-pion scattering length $a_0/m_\pi$ ($1/{\rm GeV}^2$)
obtained from the time correlation functions
and the wave functions.
}
\vspace*{-0.5cm}
\end{figure}

We try to extract the value of $k^2$ from the wave function
by fitting the data with the function defined by (\ref{Luscher.div_two}).
The results are plotted by large cross symbols 
in Figure~\ref{FIG.1.fig}, where we choose $|\vec{x}| = 7.8 - 16$ as the 
fitting range. 
The fit works very well. The result for $k^2$ is 
$2.67(12)\times 10^{-3}$ ${\rm GeV}^2$, 
which is consistent with 
that obtained from
the time dependence of the two-pion four-point function, 
$2.62(23)\times 10^{-3}$ ${\rm GeV}^2$.
We check the consistency between the two determinations 
at all simulation points on $20^3$ lattice in Figure~\ref{FIG.3.fig}, 
where the scattering length is calculated 
by substituting $k^2$ to the L\"uscher's formula (\ref{Luscher.eq}).
We observe that the statistical error of the scattering length 
obtained from the wave function is smaller than that from 
the time correlation function.
This feature was also found for the 2 dimensional XY model~\cite{Wisz.stat}.
%
%
%
%

\begin{figure}[t]
\vspace*{-0.4cm}
\centerline{\epsfxsize=7.2cm \epsfbox{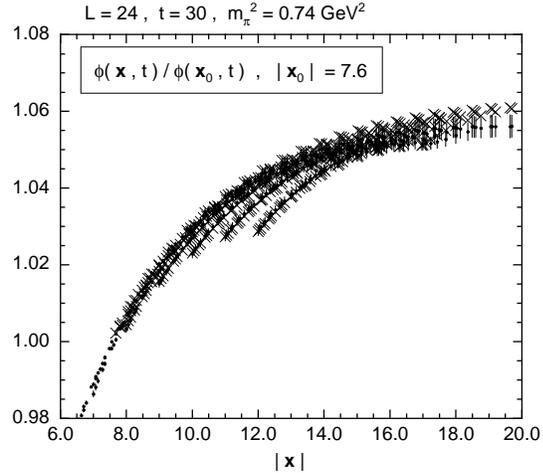}}
\vspace*{-1.0cm}
\caption{\label{FIG.4.fig}
Two-pion wave function on $24^3$ lattice at $t=30$ and $m_\pi^2=0.74 {\rm GeV}^2$.
}
\vspace*{-0.5cm}
\end{figure}

The wave function on a larger lattice of $24^3$ at $m_\pi^2=0.74{\rm GeV^2}$ 
and the results of the fitting are shown in Figure~\ref{FIG.4.fig}, 
where we choose $t=30$.
We find a systematic discrepancy between the data and the fit
for large $|\vec{x}|$ region ($|\vec{x}| > 15$).
Further there is a large discrepancy in the scattering length 
obtained from two methods as shown in Figure~\ref{FIG.3.fig}.

A possible cause of the discrepancy is smallness of the time extent.
The strength of the two-pion interaction on a finite $L^3$ periodic box 
is proportional to $1/L^3$.
If we consider a time-dependent effective Shr\"odinger equation, 
the wave function satisfies a heat diffusion equation.
We expect that a larger time extent is necessary to obtain 
the correct wave function for a larger volume.
In order to confirm this anticipation, 
we are going to explore simulations on a temporal extent $T=80$ in our 
next study.
%
%
%
%

\hfill\break
This work is supported in part by Grants-in-Aid of the Ministry of Education 
(No.15740134). 
Simulations were performed on the parallel computer PILOT3
at the Center for Computational Physics, University of Tsukuba.
%
%

%
%
\end{document}